\documentclass[a4paper,12pt]{article}
\usepackage{graphicx}
\usepackage{a4wide}
\usepackage{amsmath}
\DeclareMathOperator{\IM}{Im}
\DeclareMathOperator{\RE}{Re}
\usepackage{cleveref}
\crefdefaultlabelformat{#2\bfseries\upshape(#1)#3}
\creflabelformat{equation}{#2\bfseries\upshape(#1)#3}
\begin{document}
\title{
 \begin{flushright}
 {\small CERN-TH-2020-066}
\end{flushright}\vskip 2cm
{\bf New result on phase shift analysis}\footnote{To the memory of  Henri  Cornille, Claude Itzykson, and Joachim Kupsch}}
\author{Andr\'e Martin$^*$\\
{\small Physics Department, CERN, Geneva, Switzerland}\\
{\small $^*$e-mail: andre.martin@cern.ch}\\
{\small and}\\
Jean-Marc Richard$\dagger$\\
{\small Universit\'e de Lyon, Institut de Physique des 2 Infinis de Lyon}\\ 
{\small UCBL--IN2P3-CNRS, 4, rue Enrico Fermi, Villeurbanne, France}\\
{\small $^\dagger$e-mail: j-m.richard@ipnl.in2p3.fr}}
\date{\small \today}
\maketitle
\begin{abstract}
\noindent
Assuming a certain continuity property, we prove, using the old results of Itzykson and Martin, that, except for an obvious ambiguity, there are only at most two amplitudes reproducing an elastic differential cross section at a given energy.
\end{abstract}
\section{Historical introduction}
Given a differential cross section at one energy in the elastic region, can one find the scattering amplitudes? We restrict ourselves to the spinless case. Forgetting kinematical factors (more precisely, we neglect $1/k$ in the amplitude and $4\,\pi/k^2$ in the integrated cross-section), the scattering amplitude is given by
\begin{equation}\label{eq1}
f(\cos\theta)=\sum_\ell (2\,\ell+1)\,f_\ell\,P_\ell(\cos\theta)~,
\end{equation}
with $f_\ell=\sin\delta_\ell\,\exp(i\delta_\ell)$. The $\delta_\ell$'s are the phase shifts. The differential cross section is given by $F^2$, where
\begin{equation}
F= |f (\cos\theta )|.
\end{equation}
 It  seems that it was first realized by T Y.~Wu and T.~Ohmura in 1962 \cite{ref1} that the phase $\phi$ of the scattering amplitude  such that
\begin{equation}\label{eq3}
f (\cos \theta )= F (\cos \theta)\exp[i \phi(\cos \theta)]~,
\end{equation}
satisfies a non-linear integral equation which, using numbers to designate directions, is
\begin{equation}
F(12) \sin\phi(12)= \frac {1}{4\,\pi} \int d \Omega_3 F(13)\,F(23)\,\cos [\phi(13) - \phi(23)]~.
\end{equation}
However, they did not go very far in the possible solutions of this equation. One obvious ambiguity is that one can change the sign of all phase shifts or, equivalently change $f$ to $-f^*$, but the question is whether (\ref{eq1}) has one or more solutions. J.H.~Crichton, in 1966, pointed out that there are cases where the solution of (\ref{eq1}) is not unique \cite{ref2} and he exhibited a twofold ambiguity in the simple situation where there are only 3 partial waves, $\ell=0,\,1,\,2$. Later, in 1968, A.~Martin \cite{ref3} and R.~Newton~\cite{ref4} found a condition under which the solution of (\ref{eq1}) is unique and reached by a contraction mapping. This condition was:
\begin{equation}\label{cond3}
\sup \frac{F(13)\,F(23)}{ F(12)} <0.79~.
\end{equation}

Later A.~Gangal and J.~Kupsch \cite{ref5} succeeded in replacing $0.79$ by $0.89$. Nobody realized that condition(\ref{cond3}) implies $\sup F(11)<0.79$, and by the optical theorem,
\begin{equation}
 \sigma_{\rm tot} < .79~.
\end{equation}

What happens if this condition is violated?  One can go a little further as was done by C.~Itzykson and A.~Martin \cite{ref6} assuming that there is a finite number of partial waves and showing that if the cross section is less than $1.38$, the solution is unique. Since the maximum $\ell$  is arbitrarily large, we believe that this holds also  for an infinite  number of partial  waves. 
In fact one can go further than that and show that for a given total cross section $\sigma$,   the maximum number of solutions is less than $2^{8\,\sigma/7}$ (see appendix~\ref{app:A}).
However, after the Crichton example, it was discovered that there are other situations with 2 and only 2 solutions.  Courageously  F.~Berends and  S.~Ruijsenaars \cite{ref7} and independently  H.~Cornille \cite{ref8} proved that if $\ell=0,\,1,\,2,\,3$, there  are only 2 solutions, and H.~Cornille and J.-M.~Drouffe made the "tour de force" to prove the same for maximum $\ell=4$ \cite{ref9}.

C.~Itzykson and A.~Martin~\cite{ref6} undertook a different approach, assuming that the scattering amplitude is a genuine entire function, i.e., not a polynomial, and succeeded to prove that in most cases there are not more than 2 solutions. In this paper we shall utilise their result to go further. First we give a short lecture on entire functions in the next section, then we re-derive some of the results of \cite{ref6}. In the following section we modify a polynomial amplitude, to make it an entire function and succeed to prove that  for polynomial amplitudes of arbitrarily large degree there are only 2 solutions. Finally, we move to amplitudes with infinite number of partial waves and succeed, we believe, in proving that there are only 2 solutions. 
\section{Lecture on entire functions}
We begin by recalling some facts on entire functions which can be found in the book by Boas~\cite{ref10}. An entire function, $f(z)$ is analytic in the full complex plane. This, of course, includes polynomials. We are interested only in genuine entire functions which are such that if $M(r)$ is the maximum of $|f(z)|$ on $|z|=r$ 
\begin{equation}
\frac{\ln M(r)} {\ln r}\rightarrow \infty\quad \text{for}\quad r\rightarrow \infty~.
\end{equation}
 An entire function is of order $\rho$ if $\lim \ln M(r)/r^\rho$ is finite. For instance $\exp z$ is of order 1.
A product of 2 functions of order $\rho$ is also of order $\rho$.
 An alternative definition is obtained from a power series expansion of $f$
\begin{equation}\label{eq7}
 f=\sum_n a_n\, z^n~.
\end{equation}
Then    
\begin{equation}\label{eq8}
 \rho= \lim\sup_n \frac{n\,\ln n}{-\ln a_n }~. 
\end{equation}
An important property of functions of order strictly less than 1 is that they can be written as a convergent product over the zeros of the function
\begin{equation}\label{eq9}
f(z)=z^m\,\prod_i(1-z/z_i)~. 
\end{equation}
This will be the case of functions of order 1/2 like  
\begin{equation}
 f= \cos (\sqrt{z})=\prod_n\left[1-\frac{z}{((n+1/2)\,\pi)^2}\right]~.
\end{equation}
Let us also point out that the only function of order 1 which has no zero is $\exp(c\,z)$.

Now, we can also define entire functions as Legendre polynomial expansions.
 \begin{equation}\label{eq10}
f(z)=\sum_\ell (2\,\ell+1)\,f_\ell\,P_\ell(z)~.
 \end{equation}
The connection between polynomial expansion and  power-series expansion is rather obvious because of the inequalities
\begin{equation}\label{eq11}
|z|^\ell < |P_\ell(z)|< (1+\sqrt2)^\ell |z|^\ell~,
\end{equation}
for $|z]>1$ (see appendix \ref{app:B}).
So, again, the order of the expansion (\ref{eq10}) is given by
\begin{equation}
 \rho= \lim \sup_\ell\frac{\ell\,\ln (\ell)}{-\ln|f_\ell|}~.                                                             
\end{equation}
\section{Some results of Itzykson-Martin~\cite{ref6})}
We consider only the situation where the scattering amplitude is an entire function of order~1. It is given by \eqref{eq10}.
The unitarity condition is
\begin{equation}\label{eq:unit}
 \IM f_\ell=|f_\ell|^2~.
\end{equation}
If $f$ is entire, of order 1
\begin{equation}
 \lim\sup_\ell\frac{\ell\,\ln\ell}{-\ln |f_\ell|}=1~.
\end{equation}
Then
\begin{equation}
 \lim\sup_\ell\frac{\ell\,\ln\ell}{-\ln(\IM f_\ell)}=\frac12~.
\end{equation}
So the absorptive part
\begin{equation}
 A=\sum_\ell (2\,\ell+1)\,\IM f_\ell\, P_\ell(z)=\frac{1}{2i}\left[f(z)-f^*(z^*)\right]~,
\end{equation}
is of order 1/2, while the dispersive part
\begin{equation}
 D=\sum_\ell (2\,\ell+1)\,\RE f_\ell\, P_\ell(z)=\frac{1}{2}\left[f(z)+f^*(z^*)\right]~,
\end{equation}
is of order 1.

Suppose we have 2 amplitudes with the same differential cross section:
\begin{equation}
f= D+i A~, \quad f'= D'+i A'~, \quad |f|^2=|f'|^2~,
\end{equation}
then
\begin{equation}
D^2-D'^2=A'^2-A^2=Q~.
\end{equation}
$Q$ is a function of order 1/2 and, therefore, can be written as a convergent product
\begin{equation}
Q=z^p\prod_i\left(1-z/z_i\right)~.
\end{equation}
But 
\begin{equation}
 -Q= (D'-D)( D'-D)~.
\end{equation}
The zeros of  $D'-D$ and $D'+D$  form two complementary subsets of the zeros of $Q$. So
\begin{equation}
D'-D=M(z)\,z^m\,\prod_\alpha (1-z/z_\alpha)~,
\end{equation}
where $M(z)$ has no zero and the product is still convergent and of order 1/2.
Similarly
\begin{equation}
D'+D=N(z)\,z^n\,\prod_\beta (1-z/z_\beta)~,
\end{equation}
where $N(z)$ has no zero. $M$ and $N$ being of order 1 without zeros can only be of the form $\exp(c\,z),\, \exp(c'\,z)$, but since they disppear in the product, $c'=-c$. So
\begin{equation}\label{eq:decomp}
  2\,D= \exp(c\,z)\, m(z) + \exp(-c\,z)\, n(z)~,
\end{equation}
where $m(z)$ and $n(z)$ are of order 1/2.

Suppose now that we have a third amplitude, $f"= D"+iA"$.
Then $D$ and $D"$ will have a different decomposition:
 \begin{equation}
      2\,D=\exp(d\,z)\, p(z)+ \exp(-d\,z)\, q(z)= \exp(c\,z)\, m(z) + \exp(-c\,z)\, n(z)~.
\end{equation}
Suppose   $c>d>0$. Then the term containing  $\exp(c\,z)$ dominates. The second equality is impossible and necessarily, $c=d$. Then we have
 \begin{equation}
\exp(2\,c\,z)\, ( m-p) + n-q =0~,
\end{equation}
but $m$, $p$, $n$, and $q$ being of order 1/2 this is impossible. So the decomposition  (\ref{eq:decomp}) is unique, and there are at most two solutions.
\section{Ambiguities for polynomial amplitudes}
We consider an amplitude
\begin{equation}
 \mathcal{F}_L=\sum_{\ell=0}^L (2\,\ell+1)\,f_\ell\,P_\ell(\cos\theta)=D_L+i A_L~,\quad \text{with} \quad \IM f_\ell=|f_\ell|^2~,
\end{equation}
where $L$ is arbitrary. We shall complete this amplitude by a unitary amplitude extending from $L+1$ to infinity and manufacture in this way an entire function of order 1. We add to $\mathcal{F}_L$:
\begin{equation}
 R_L(\lambda)=\sum_{L+1}^\infty (2\,\ell+1)\,r_\ell(\lambda)\,P_\ell(\cos\theta)~,\quad\text{with}\quad
 r_\ell(\lambda)=\RE r_\ell+ i\IM r_\ell~,
\end{equation}
and we take, with $|\lambda|<1/2$,
\begin{equation}
 \RE r_\ell=\frac{\lambda}{2}\int_{-1}^{+1} P_\ell(x) \,\exp x \,dx~.
\end{equation}
Though this can be calculated explictely term by term, we need only an asymptotic estimate for large $\ell$ (see Appendix \ref{app:B}).
\begin{equation}
 \RE r_\ell\simeq \frac12\,\frac{1}{\ell!}\frac{\lambda}{2^\ell}\,\frac{\sqrt{\pi}}{\sqrt{\ell-1/2}}~.
\end{equation}
The dispersive part of the new amplitude is:
\begin{equation}
 D_L+\sum_{L+1}^\infty (2\,\ell+1)\,\RE r_\ell\, P_\ell(z)~.
\end{equation}
It is an entire function of order 1. The absorptive part is constructed to satisfy elastic unitarity for all $\ell$'s. So, for $\ell\le L+1$,
\begin{equation}
 \IM r_\ell(\lambda)=\dfrac{1-\sqrt{1-4 (\RE r_\ell)^2}}{2}~.
\end{equation}
Notice that as soon as
\begin{equation}
 \RE r_\ell<\frac25~,
\end{equation}
we have
\begin{equation}
 \IM r_\ell<\frac54 |\RE r_\ell|^2~.
\end{equation}
So, asymptotically
\begin{equation}
 \IM r_\ell<\frac54\left(\frac12\,\frac{1}{\ell!}\,\frac{1}{2\,\ell}\right)^2\,\frac{\lambda^2\,\pi}{\ell-1/2}~.
\end{equation}
This means that the new absorptive part of the amplitude
\begin{equation}
 A_L+\sum_{L+1}^\infty (2\,\ell+1)\,\IM r_\ell\, P_\ell(z)~,
\end{equation}
is an entire function of order 1/2. So the Itzykson-Martin theorem applies to this amplitude, and there cannot be more than one amplitude giving the same differential cross section. The differential cross-section  is a continuous function of $\lambda$, including $\lambda= 0$. it is even real analytic in a neighbourhood of   $\lambda= 0$.
We do not see how extra amplitudes (in finite number from Appendix~\ref{app:A}) could appear at $\lambda= 0$. If they did, they should also be present for $\lambda$ different from zero. This is a continuity assumption, which seems to us reasonable, but which could be criticized.            .
\section{Ambiguities for non polynomial amplitudes}
Suppose we have an amplitude which is neither a polynomial nor an entire function. This amplitude could exist on the interval $[-1,\,+1]$  or be analytic in the complex plane with some singularities: Its partial wave expansion will converge in an ellipse with foci $\{-1, +1\}$ touching the nearest singularity. This is the case in local field theory \cite{ref9}. Then  with   $f_\ell=\sin \delta_\ell\,\exp(i\delta_\ell)$,  
\begin{equation}\label{eq33}
  \RE f_\ell   < \exp(-k\,\ell)~,\quad\text{and} \quad  
    \IM f_\ell    < \exp(- 2\,k\,\ell) ~.               
\end{equation}

Now we replace $\delta_\ell$ by   $\delta_\ell\,\exp(-\lambda\,\ell\,\ln\ell)$ and   correspondingly    $f_\ell$       by    $f_\ell(\lambda)$.
The new dispersive part $D(\lambda)$ is an entire function of order 1 and the  absoptive part $A(\lambda)$ an entire function of order 1/2. So with this amplitude there is at most a two-fold ambiguity. This persists as  $\lambda$   goes to zero and we do not see how extra solutions could appear out of nowhere. This is an assumption which  seems reasonable.
\section{Concluding remarks}
 We consider that we have solved a long standing problem. However, purists could object that we rely on a limiting  process and that we don't know if a catastrophe could occur in the limit. 
We don't believe this because the differential cross section remains continuous at the limit. 
We apply ``Goldberger's principle'' that ``nature cannot be so unkind'' \cite{ref12}. 
We hope that someone else can make our argument more rigorous. However there is a big problem left which is the existence of solutions. For physical  cross sections like $\pi\pi$ scattering, physics tells us that the solution exists, but assuming an arbitrary differential cross section is there at least one acceptable amplitude? 
It is only under a condition similar to \eqref{cond3}, where $0.79$ is replaced by 1, that we are certain of the existence of a solution.
\subsection*{Acknowledgments}
We would like to thank Shirin Davarpanah for her help in preparing the manuscript.
\appendix
\numberwithin{equation}{section}
\section{Bound on the number of amplitudes}\label{app:A}
In Ref.~\cite{ref6}, we have shown that if  the scattering amplitude is a polynomial of degree $L$, and, therefore, the differential cross-section  a polynomial of degree $2\,L$, the scattering amplitude is unique if the cross section is less than 1.38. It is tempting to assume that this result, independent of $L$ is also valid for an infinite number of partial waves. Here, we want to generalize this result and prove that, given the total cross section, one can find a bound on the maximum number of amplitudes  compatible with the differential cross section which is \emph{independent} of $L$ and depends only on the value of the total cross-section, while, naively one would expect a bound of $2^{L-1}$. The method is the same as in \cite{ref6}. It consists in  starting with the maximum $\ell$, i.e., $L$, and descending in values of $\ell$. From the Legendre polynomial expansion of the differential cross.section:
\begin{figure}[t]
\centering
\includegraphics[width=.55\textwidth]{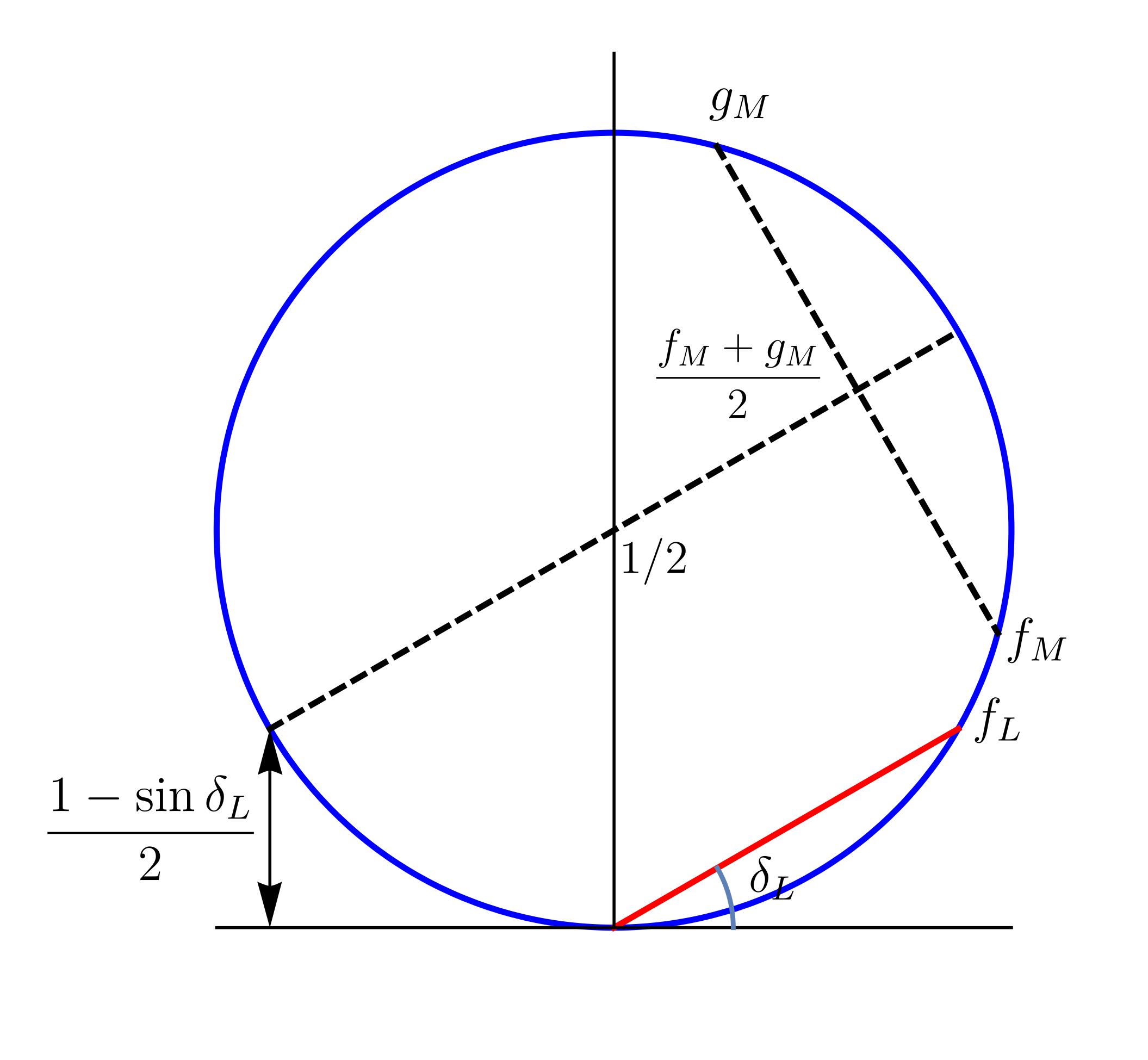}
 \caption{Notations used in appendix \ref{app:A}}
 \label{fig:fig1}
\end{figure}
\begin{equation}
\sum_0^{2\,L} (2\,\ell+1)\,C_\ell\, P_\ell(\cos\theta)~, 
\end{equation}
we see  that the value of $\sin(\delta_L)$ is fixed by the knowledge of $C_{2L}$ (we choose $\delta$ between 0 and $\pi/2$). Then, from $C_{2L-1}$, we get
\begin{equation}
 \RE( f_{L-1}\, f_L^*)~.
\end{equation}
This defines in the Argand diagram a straight line which intersects the unitarity circle in 2 points (if they don't intersect, the differential cross section is not acceptable). If only one of these 2 points is acceptable (and we shall see that soon) we  can continue the procedure down to a value $\ell=M$. Suppose now that at $\ell=M$ we have the 2 solutions, $f_M$ and $g_M$. Then we have
\begin{equation}
 \RE( f_{M}-g_M) f_L^*=0~.
\end{equation}
So $f_M-g_M$ is perpendicular to the segment
$[0,f_L]$. The point $(f_M+g_M)/ $2 is on a line parallel to $[0,f_L]$ going through the centre of the unitarity circle.  See Fig.~\ref{fig:fig1}. It is easy to see that since the segment $[f_M,g_M]$ must at least touch the circle we must have
 \begin{equation}
  \frac{\IM f_M+\IM g_M}{2}> \frac{1-\sin\delta_L}{2}~.
 \end{equation}
 So the total cross section, common to both amplitudes, should be larger than
\begin{equation}
(2\,M+1)\,\frac{1-\sin\delta_L}{2}+ (2\,L+1)\,\sin^2 \delta_L~.
\end{equation}
Minimizing with respect to $\delta_L$, and noticing that $M<L$ we get
\begin{equation}
\sigma_\text{tot}> (7/8)(M+1/2)~.
\end{equation}
 If, on the other hand,
\begin{equation}
 (7/8) (M+3/2)>\sigma_\text{tot}> (7/8)(M+1/2)~,
 \end{equation}
we see that there cannot be any ambiguity  for $\ell\ge  M+1$, which we anticipated, and we must choose the solution with the smallest imaginary part.

At this stage we can have ambiguities at each step descending to $\ell=1$. As $\ell=0$ is fixed, an upper bound on the number of solutions is
\begin{equation}
  2^{(M-1)} <  2^{7/(8\,\sigma_\text{tot})}.
\end{equation}
This bound is independent of $L$. We believe that this also holds for an infinite number of partial waves. What really matters for the rest of this paper is that the number of solutions is finite.
\boldmath\section{Proof of the inequalities  \cref{eq11}}      
\unboldmath \label{app:B}
1) $|P_\ell (z)|> P_\ell (|z|$ for   $|z|>1$. The Legendre polynomials have zeros between $-1$ and $+1$. Hence, taking for instance $\ell$ even
\begin{equation}
 |P_\ell(z)|=C_\ell\,\left|\prod_i(z^2-x_i^2)\right|>
 C_\ell\,\prod_i(|z|^2-x_i^2)=P_\ell(|z|)~.
\end{equation}
Now, $P (x)$ for $x$ real larger than 1 can be written as
\begin{equation}
 P_\ell(x)=\frac1\pi\int_0^{\pi/2}
 \left[(x+\cos\phi\,\sqrt{x^2-1})^\ell+
       (x-\cos\phi\,\sqrt{x^2-1})^\ell\right]\,d\phi>x^\ell~.
\end{equation}
This proves the left hand inequality. The second inequality follows from the above integral representation: for $|z|\ge1$,
\begin{equation}
 |P_\ell(z)|<\frac{|z|^\ell}{\pi}\,\int_0^\pi
\left(1+\sqrt{1+1/|z|^2}\right)^\ell\,d\phi<|z|^\ell\,(1+\sqrt2)^\ell~.
\end{equation}
\boldmath\section{Asymptotic behaviour of $\RE r_\ell$ for large $\ell$}\unboldmath
We have 
\begin{equation}\label{eq:r-ell}
 \RE r_\ell=\frac\lambda2\,\int_{-1}^{+1}P_\ell(x) \exp x \, dx~.
\end{equation}
From the definition of Legendre polynomials
\begin{equation}\label{eq:Leg-def}
 P_\ell(x)=\frac{1}{2^\ell\, \ell!}\genfrac{(}{)}{}{}{d}{dx}^\ell\,(x^2-1)^\ell~,
\end{equation}
we can integrate the RHS of \eqref{eq:r-ell} $\ell$ times by parts: 
\begin{equation}
 \RE r_\ell =\frac{\lambda}{2^\ell\, \ell!}\,
 \int_{-1}^{+1} \cosh x\, (1-x^2)^\ell\,dx~.
\end{equation}
This gives for large $\ell$
 \begin{equation}
 \RE r_\ell\simeq \frac\lambda2\,\frac{1}{2^\ell\,\ell!}\int_{-\infty}^{+\infty}\exp[-(\ell-1/2) x^2] \, dx=
  \frac\lambda2\,\frac{1}{2^\ell\,\ell!}\,\sqrt{\frac{\pi}{\ell-1/2}}~.
 \end{equation}
\end{document}